\documentclass[conference]{IEEEtran}
\IEEEoverridecommandlockouts

\usepackage{cite}
\ifCLASSINFOpdf
\else
\fi
\usepackage{amsmath}
\makeatletter
\newcommand{\pushright}[1]{\ifmeasuring@#1\else\omit\hfill$\displaystyle#1$\fi\ignorespaces}
\newcommand{\pushleft}[1]{\ifmeasuring@#1\else\omit$\displaystyle#1$\hfill\fi\ignorespaces}
\makeatother

\usepackage{graphicx}

\usepackage{url}
\usepackage{color,soul}

\usepackage{textcomp}

\begin{document}
%
% paper title
% Titles are generally capitalized except for words such as a, an, and, as,
% at, but, by, for, in, nor, of, on, or, the, to and up, which are usually
% not capitalized unless they are the first or last word of the title.
% Linebreaks \\ can be used within to get better formatting as desired.
% Do not put math or special symbols in the title.
%\title{Case Study Analysis of Degradation Cost's Effects on the Operation of a Price-maker Battery Energy Storage System in the Electricity Markets}

\title{Impact of Battery Degradation on Market Participation of Utility-Scale Batteries: Case Studies}

% author names and affiliations
% use a multiple column layout for up to three different
% affiliations
\author{
\IEEEauthorblockN{Reza Khalilisenobari and Meng Wu}
\IEEEauthorblockA{School of Electrical, Computer and Energy Engineering, Arizona State University\\
Tempe, Arizona \\
Email: rezakhalili@asu.edu - mwu@asu.edu}

\thanks{This work was supported by the Power Systems Engineering Research Center.}
}

% conference papers do not typically use \thanks and this command
% is locked out in conference mode. If really needed, such as for
% the acknowledgment of grants, issue a \IEEEoverridecommandlockouts
% after \documentclass

% for over three affiliations, or if they all won't fit within the width
% of the page, use this alternative format:
% 
%\author{\IEEEauthorblockN{Michael Shell\IEEEauthorrefmark{1},
%Homer Simpson\IEEEauthorrefmark{2},
%James Kirk\IEEEauthorrefmark{3}, 
%Montgomery Scott\IEEEauthorrefmark{3} and
%Eldon Tyrell\IEEEauthorrefmark{4}}
%\IEEEauthorblockA{\IEEEauthorrefmark{1}School of Electrical and Computer Engineering\\
%Georgia Institute of Technology,
%Atlanta, Georgia 30332--0250\\ Email: see http://www.michaelshell.org/contact.html}
%\IEEEauthorblockA{\IEEEauthorrefmark{2}Twentieth Century Fox, Springfield, USA\\
%Email: homer@thesimpsons.com}
%\IEEEauthorblockA{\IEEEauthorrefmark{3}Starfleet Academy, San Francisco, California 96678-2391\\
%Telephone: (800) 555--1212, Fax: (888) 555--1212}
%\IEEEauthorblockA{\IEEEauthorrefmark{4}Tyrell Inc., 123 Replicant Street, Los Angeles, California 90210--4321}}

% use for special paper notices
%\IEEEspecialpapernotice{(Invited Paper)}

\IEEEpubid{\makebox[\columnwidth]{978-1-7281-8192-9/21/\$31.00~\copyright2021 IEEE \hfill} \hspace{\columnsep}\makebox[\columnwidth]{ }}

% make the title area
\maketitle

% As a general rule, do not put math, special symbols or citations
% in the abstract
\begin{abstract}
The recent decrease in battery manufacturing costs stimulates the market participation of utility-scale battery energy storage systems (BESSs). However, battery degradation remains a major concern for BESS owners while determining their BESS investment and operation strategies. This paper studies the impact of battery degradation on BESS's operation and revenue/cost in real-time energy, reserve, and pay as performance regulation markets. Comparative case studies are performed on two optimization frameworks which model the participation of a price-maker BESS in energy and ancillary services markets with and without considering battery degradation cost. A synthetic test system built upon real-world data is adopted in the case studies. Analyses reveal that degradation cost plays an important role in the scheduling of BESSs and should not be neglected. Several potential enhancements to the optimization frameworks are discussed based on the performed analyses. 

\end{abstract}
% no keywords

% For peer review papers, you can put extra information on the cover
% page as needed:
% \ifCLASSOPTIONpeerreview
% \begin{center} \bfseries EDICS Category: 3-BBND \end{center}
% \fi
%
% For peerreview papers, this IEEEtran command inserts a page break and
% creates the second title. It will be ignored for other modes.
\IEEEpeerreviewmaketitle

\section{Introduction}

The grid integration of battery energy storage systems (BESSs) is expanding rapidly, thanks to the BESS's desirable characteristics of being a fast, efficient, and flexible generating resource with the capability of multiple services provision \cite{BESSapp}. BESSs are capable of providing a wide range of grid services including energy arbitrage, frequency regulation, reserve provision, resiliency enhancement, and renewable firming \cite{market_trend,ref40}. In recent years, the battery manufacturing cost has reduced, which encourages merchant BESSs to participate in the electricity markets. Since 2018, independent system operators (ISOs) in the US are required by the Federal Energy Regulatory Commission (FERC) Order 841 to remove barriers for BESS's participation in the various markets \cite{Order841}. This order further motivates merchant BESSs to offer energy and ancillary services through the wholesale market. It also inspires the power industry and research community to thoroughly understand the BESS's operating patterns, revenue streams, and cost structures in the energy and ancillary services markets.

To study the market operation of merchant BESSs, existing literature models the merchant BESS as a price-maker in the market, since these utility-scale BESSs could impact market prices due to their capacity and technical characteristics \cite{ref25,ref22,ref24,MvsT2,NAPS,arxiv}. Using the price-maker modeling approach, it is possible to study the optimal allocation of merchant BESS across various markets and analyze the impact of BESS's profit maximization activities on the ISO's market clearing outcomes.

In line with the works which focus on markets' design\cite{arxiv2} and operation \cite{OPR}, reference \cite{ref25} compares three different market mechanisms for operation scheduling of a BESS in the energy market. This work is mainly focused on the market mechanism and does not deal with operational details of BESS and ancillary services markets. A comprehensive study on the operation of price-maker BESS in the energy market under various transmission congestion scenarios is performed in \cite{ref22}. This work does not model ancillary services markets besides the energy market and also neglects the battery degradation cost. Studies on market operations of price-maker BESSs are enhanced in \cite{ref24} by considering day-ahead energy and reserve markets and real-time balancing market. However, the frequency regulation market, which is one of the major revenue streams for BESSs in practical applications \cite{market_trend}, is not considered in \cite{ref24}. In \cite{MvsT2}, BESS is modeled as a price-taker in the energy market and price-maker in reserve and frequency regulation markets using a novel approach for price-maker modeling. Reference \cite{MvsT2} models the most important markets for the operation of BESS and also considers parameter uncertainties in the model. However, this work neglects the battery degradation cost, which may lead to inaccurate results. 

To enable comprehensive analysis on the interactions between merchant BESSs and energy/ancillary services markets, we propose two optimization frameworks in \cite{NAPS,arxiv} for the participation of price-maker BESS in real-time energy, reserve, and pay as performance frequency regulation markets with (in \cite{arxiv}) and without (in \cite{NAPS}) an accurate degradation cost model. In \cite{arxiv}, we also develop a market-based dispatch model for the automatic generation control (AGC) signals, in order to study the details of BESS operations in the regulation market.

Built upon our frameworks in \cite{NAPS,arxiv}, this paper performs comparative case studies to investigate the impact of battery degradation on the operation scheduling and revenue/cost of a price-maker BESS in real-time energy, reserve, and pay as performance regulation markets. The frameworks in \cite{NAPS,arxiv} enable us to determine the optimal scheduling of a price-maker BESS is real-time energy and ancillary services markets with and without considering the battery degradation cost in the decision-making process. Simulation results obtained using each framework on a synthetic test system are presented in four different cases, representing various market participation policies for the BESS. A comparative analysis of the case study results evaluates the impact of battery degradation on the BESS's market operation.

In the remainder of the paper, Section II provides a general overview of the two optimization frameworks with \cite{arxiv} and without \cite{NAPS} the degradation cost model. Section III compares the case study results and discusses the impact of battery degradation on BESS's participation in real-time energy, reserve, and regulation markets. Section IV concludes the paper.

\section{Frameworks Description}

In our previous works \cite{NAPS,arxiv}, we propose two optimization frameworks to study the optimal participation of a price-maker BESS in real-time energy, reserve, and pay as performance frequency regulation markets. Both frameworks contain comprehensive models for BESS operating limits, as well as details of ISO's joint market clearing procedure for real-time energy and ancillary services markets. The framework in \cite{NAPS} neglects battery degradation costs for simplicity, while the framework in \cite{arxiv} contains an accurate degradation cost model. 

To study detailed interactions between BESS's profit maximization strategies and ISO's joint market clearing procedure across real-time energy, reserve, and regulation markets, in \cite{NAPS,arxiv}, the BESS is modeled as a price-maker using bi-level optimization frameworks with coupled upper and lower level problems. In the upper-level problem (ULP), BESS's owner maximizes its revenue across various markets within a certain time period, while the lower-level problem (LLP) simulates the ISO's joint market clearing procedure. BESS's quantity and price offers for each market at each market clearing interval serve as decision variables in the ULP and input parameters to the LLP, while market clearing prices (MCPs) and scheduled power of BESS serve as decision variables in the LLP and input parameters to the ULP.

\begin{figure}[!h]
	\centering
	\includegraphics[width=3.1in, height=3.59in]{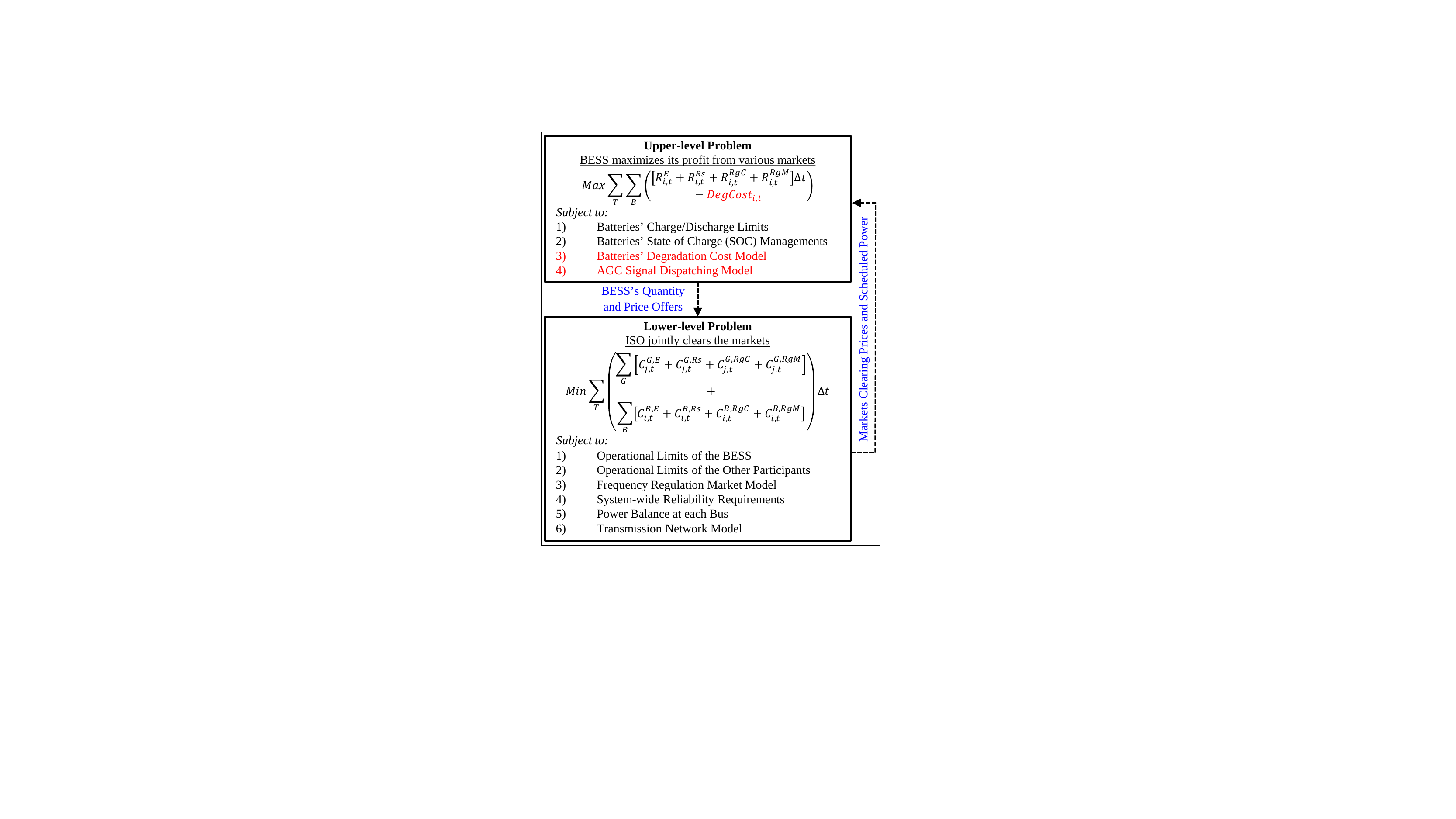}
	\caption{Structure of the optimization frameworks. $T$ denotes the studied time period; $B$ denotes the set of all battery units owned by a BESS owner; $G$ denotes the set of all the other generating units in the market; $\Delta t$ denotes the length of each market clearing interval; $R_{i,t}^{E}, R_{i,t}^{Rs}, R_{i,t}^{RgC}, R_{i,t}^{RgM}$ denote the BESS's revenue obtained using battery unit $i$ from energy, reserve, regulation capacity and regulation mileage markets at market clearing interval $t$, respectively; $DegCost_{i,t}$ denotes the degradation cost of battery unit $i$ during market clearing interval $t$; $C_{i,t}^{B,E}, C_{i,t}^{B,Rs}, C_{i,t}^{B,RgC}, C_{i,t}^{B,RgM}$ denote the operating costs of battery unit $i$ in the energy, reserve, regulation capacity and regulation mileage markets at market clearing interval $t$, respectively; $C_{j,t}^{G,E}, C_{j,t}^{G,Rs}, C_{j,t}^{G,RgC}, C_{j,t}^{G,RgM}$ denote the operating costs of generating unit $j$ in the energy, reserve, regulation capacity and regulation mileage markets at market clearing interval $t$, respectively.}
	\label{strucure}
\end{figure}

Fig. \ref{strucure} shows the structure and main components of each framework with \cite{arxiv} or without \cite{NAPS} considering battery degradation cost. The coupling between ULP and LLP is represented by terms in blue. Terms in black represent components that are modeled in both \cite{NAPS} and \cite{arxiv}. Terms in red represent components that are modeled in \cite{arxiv} only. As shown in Fig. \ref{strucure}, these two frameworks share the same LLP but differ in the ULP. A general description of the ULPs and the common LLP for both frameworks is presented below. Detailed equations for the frameworks with and without modeling battery degradation cost can be found in \cite{arxiv} and \cite{NAPS}, respectively.

\subsection{Upper-level Problem (ULP)}

In the ULP, the BESS owner who has several battery storage units at different buses maximizes its revenue from the real-time energy, spinning reserve, and pay as performance frequency regulation markets while considering its operating limits, including or not including the battery degradation cost. The ULP of each framework is described below.

\subsubsection{Framework without Degradation Cost Model}\hfill \par

In this framework, components of the ULP are represented by terms in black in Fig. \ref{strucure}. The objective function maximizes BESS's total revenue from energy, reserve, regulation capacity, and regulation mileage markets across a certain time period and across all the battery units at different buses. Each revenue term is calculated by multiplying the battery unit's dispatched power and corresponding MCP. Revenue from the energy market can be positive or negative in each market clearing interval as BESS may sell (i.e., discharge) or buy (i.e., charge) energy in each interval. Revenue from the reserve market is paid to each unit for reserving its output for contingency conditions. Revenue from the regulation market consists of regulation capacity payment, which is paid based on the unit's reserved capacity for regulation services provision; and regulation mileage payment, which is paid based on the unit's contribution toward following system-level AGC signals.

Reserve services deployment is modeled in neither of the frameworks as it is related to contingency analysis. In this framework (without degradation cost model), AGC signal deployment is modeled by assuming that AGC signals have zero-mean over each market clearing interval. These zero-mean AGC signals are provided by ISOs like PJM and ISO New England (ISO-NE) \cite{ref27}. This assumption indicates following AGC signals 1) will not change the battery unit's state of charge (SOC) across each market clearing interval while charge/discharge efficiency is ideal; 2) will not cause additional cost to BESS owners if battery degradation is neglected. Therefore, a dispatch model for AGC signals is not needed when battery degradation cost is not considered.

The constraints of the ULP in this framework include: 1) Batteries' charge/discharge limits: these constraints ensure the quantity offer of BESS for each market is within the battery units' charge/discharge limits. Also, the accumulated dispatched power for various services in each market clearing interval should not violate the charge/discharge limits of the batteries. 2) Batteries' SOC management: these constraints keep track of the changes in the battery unit's SOC in each market clearing interval, which are resulted from BESS's participation in various markets. These constraints also ensure the SOC lies within its upper/lower limits and force BESS to have the same SOC at the beginning and end of each day.

\subsubsection{Framework with Degradation Cost Model}\hfill \par

In Fig. \ref{strucure}, terms in red represent components that are added to the ULP of the previous framework to build the framework considering battery degradation costs. By subtracting the degradation cost term from the objective function and adding a group of constraints for modeling degradation costs, the battery degradation costs are calculated based on a linear approximation of the rainflow algorithm \cite{ref9}. Rainflow algorithm is an accurate method for battery degradation cost modeling, which counts the number of charge/discharge cycles in the battery's operation period and assigns costs to them based on the depth of the cycles. It means that having charge/discharge cycles with lower depth results in less degradation cost. 

To obtain degradation cost during BESS's AGC signal following activities, each battery unit's contribution toward system-level AGC signals needs to be modeled. This is handled by the AGC signal dispatching model in the constraints of the ULP, where a participation factor is defined to accurately dispatch system-level AGC signals based on market outcomes. 

\subsection{Lower-level Problem (LLP)}

As shown in Fig. \ref{strucure}, the LLP models ISOs' joint market clearing process, and it is similar in both frameworks. The LLP objective function minimizes the total operating cost of real-time energy, reserve, regulation capacity, and regulation mileage markets across a certain time period, considering the operating costs of the BESS and other non-battery market participants. Each cost term is calculated by multiplying each unit's price offer to its scheduled power. The BESS's price offers are input parameters from the ULP. It is assumed that BESS can perfectly predict price offers of other participants.

The following six groups of constraints are considered for the LLP, as shown in Fig. \ref{strucure}. 1) Operational limits of BESS: these constraints limit each battery unit's scheduled power for each market below its corresponding quantity offer. 2) Operational limits of other participants: minimum and maximum generation limits of other participants are maintained in these constraints along with ramping limits. 3) Frequency regulation market model: these constraints model the impact of each unit's historical performance in regulation services provision on its current dispatch in the regulation market. 4) System-wide reliability requirements: these constraints ensure that, during each market clearing interval, the system ancillary services requirements are satisfied by the scheduled power of BESS and other market participants. The corresponding dual variables are MCPs for reserve, regulation capacity, and regulation mileage provisions. 5) Power balance at each bus: these constraints enforce the Kirchhoff's current law at each bus. The corresponding dual variables are locational marginal prices (LMPs) at various buses. 6) Transmission network model: these constraints calculate transmission line power flow and enforce line thermal limits.

Using the conversion procedure in \cite{ref28}, both bi-level frameworks are converted to mixed integer linear programming (MILP) to be solved using available commercial solvers.

\section{Case Studies and Comparative Analysis}

The above two optimization frameworks enable us to evaluate the impact of battery degradation cost on BESS's operations in various markets. This section investigates such impact through four comparative case studies performed on the above optimization frameworks under different market participation scenarios of BESS. In \cite{NAPS,arxiv}, several case studies are performed to evaluate each proposed framework's performance separately. This section focuses on understanding how battery degradation cost could affect BESS's market participation and profit maximization activities, through comparing simulation results obtained using both frameworks.

\subsection{The Test System}

To have a synthetic test system built upon real-world data, the IEEE reliability test system (RTS), updated by Grid Modernization Laboratory Consortium, i.e., RTS-GLMC, is adopted for the case studies \cite{RTS}. The simulations are performed on the third area of the RTS-GMLC network. This test system consists of 25 buses, 39 transmission lines, 26 generators, and a battery unit on Bus 13. The simulation horizon is 24 hours with 15-minute market clearing intervals (i.e., 96 market clearing intervals in total). The AGC signals are dispatched every 20 seconds for the framework with the degradation cost model and AGC signal dispatch model. 

System load, reserve, and regulation capacity requirements for the simulation horizon are determined by averaging the summer (June to August) load and ancillary services requirements of the third area in the RTS-GMLC. During the 24-hour simulation horizon, the system load varies between 1285 MW and 2345 MW. The system peak and valley loads happen in Hours 11{\texttildelow}18 and Hours 1{\texttildelow}6, respectively. Hence, it is expected that the energy market price reaches its highest and lowest values in respectively Hours 11{\texttildelow}18 and Hours 1{\texttildelow}6.

In each market clearing interval, the system's regulation mileage requirement is set to be 1.5 times the regulation capacity requirement. The sample AGC set points, provided by ISO-NE, are modified for AGC signal modeling \cite{AGC}. The modified AGC signals have zero-mean over each 15-minute market clearing interval. In each market clearing interval, the total variation of the modified AGC signals is equal to the corresponding system regulation mileage requirement. 

It is assumed, in each market clearing interval, each generator's price offer for the energy market is equal to the generation cost. Generator price offers for reserve, regulation capacity, and regulation mileage markets are 0.15, 0.4, and 0.07 times their energy price offers, respectively. The multipliers used for calculating ancillary services' price offers are derived from PJM historical data by averaging over the ratio of each ancillary service price to the energy price \cite{PJM}.

\begin{table}[!t]
	\caption{Operational information of The Battery unit}
	\label{T1}
	\centering
	\begin{tabular}{cccccc}
		\hline
		(Dis)charge & Storage & Min & Max & Initial & (Dis)charge\\
		limit & Capacity & SOC & SOC & SOC & Efficiency\\
		(MW) & (MWh) & (MWh) & (MWh) & (MWh) & ($\%$)  \\
		\hline
		50 & 200 & 20 & 180 & 90 & 95 \\
		\hline
	\end{tabular}
\end{table}

The BESS owner has a lithium-ion battery unit at Bus 13 with the operational parameters specified in Table \ref{T1}. The useful life of the battery is 6000 full charge/discharge cycles with cycle depth of 80\%, and the battery replacement cost is 200 k\$/MWh price. This information is used for incorporating the battery degradation cost into the framework with the degradation cost model. Details on the degradation cost function can be found in \cite{arxiv}.

\subsection{Case 1: Energy Market}

\begin{figure}[!h]
	\centering
	\includegraphics[width=3in, height=2.755in]{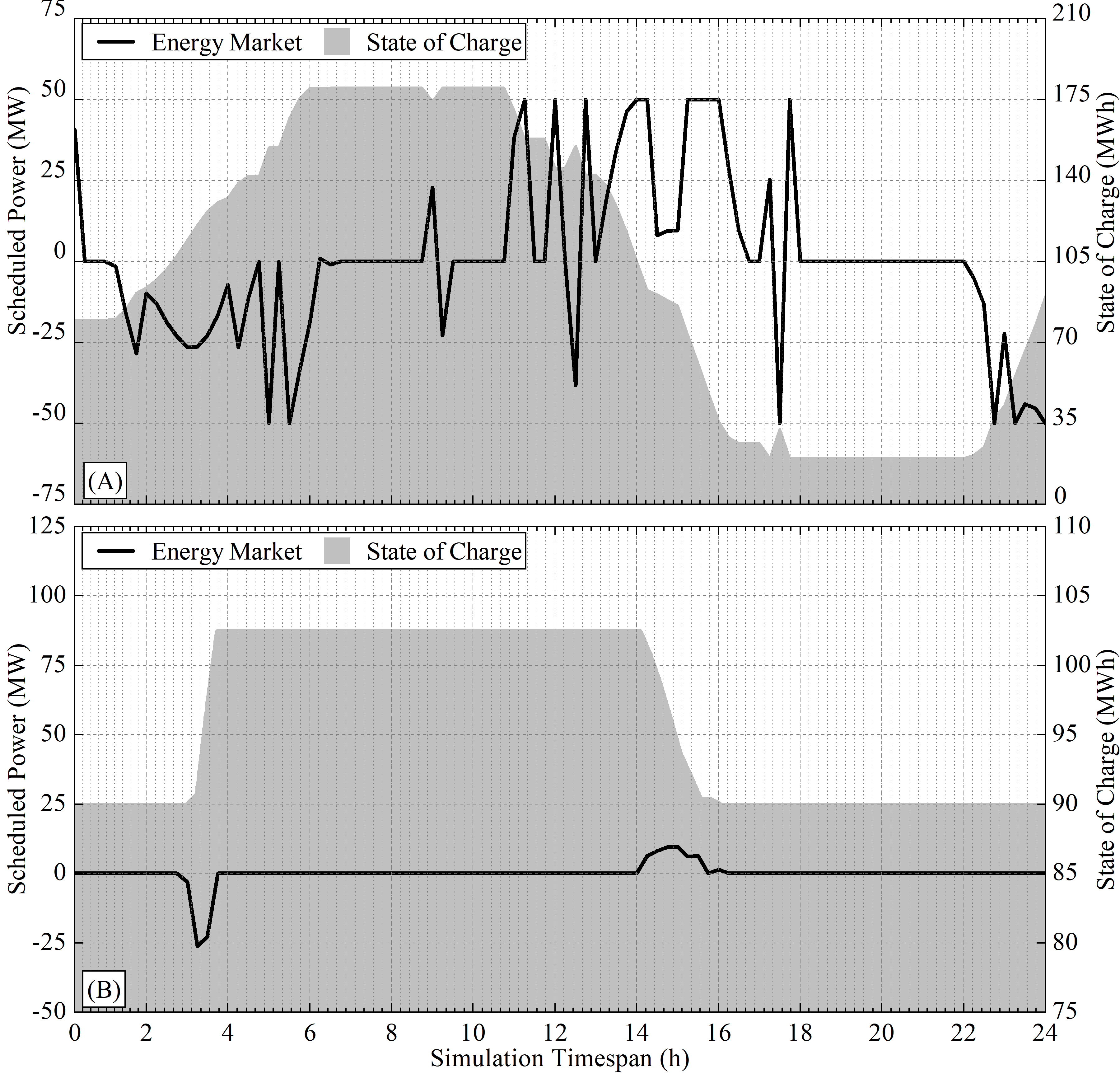}
	\caption{Scheduled power and SOC of BESS when it participates in energy market only, using (A) framework without degradation cost model; (B) framework with degradation cost model. The black curve denotes BESS's scheduled power in energy market; The grey area denotes the BESS's SOC.}
	\label{C1}
\end{figure}

This case compares the BESS's performance with and without modeling its degradation cost, when the BESS only participates in the energy market. Figs. \ref{C1}(A) and \ref{C1}(B) show the BESS's scheduled power (in MW) and SOC (in MWh) when the BESS participates in the energy market only, without and with the degradation model, respectively.

Fig. \ref{C1}(A) shows when degradation costs are neglected, energy arbitrage between hours is a profitable bidding strategy for a BESS only participating in the energy market. Hence, the BESS buys energy (with negative scheduled power) and charges itself (with an increase in the SOC) during initial hours with lower energy prices (hours 1{\texttildelow}6), in order to sell the stored energy during hours with higher energy prices (hours 11{\texttildelow}18). As shown in Fig. \ref{C1}(A), BESS uses all its available capacity for this inter-temporal energy arbitrage. The BESS also experiences deep charge/discharge cycles for maximizing its revenue, which, in real-world practices, may result in considerable battery life loss and degradation cost. 

Fig. \ref{C1}(B) shows when degradation costs are considered in the BESS's bidding strategies, the BESS's participation in the energy market is very limited. It only buys 12 MWh energy once during off-peak hours (hours 3{\texttildelow}4) and sells this amount of energy once during peak hours (hours 14{\texttildelow}16). In this case, when battery degradation is considered, participating in the energy market does not generate enough revenue to overcome the total degradation cost. Therefore, the BESS limits its participation in the energy market.

\subsection{Case 2: Energy and Reserve Markets}

\begin{figure}[!h]
	\centering
	\includegraphics[width=3in, height=2.755in]{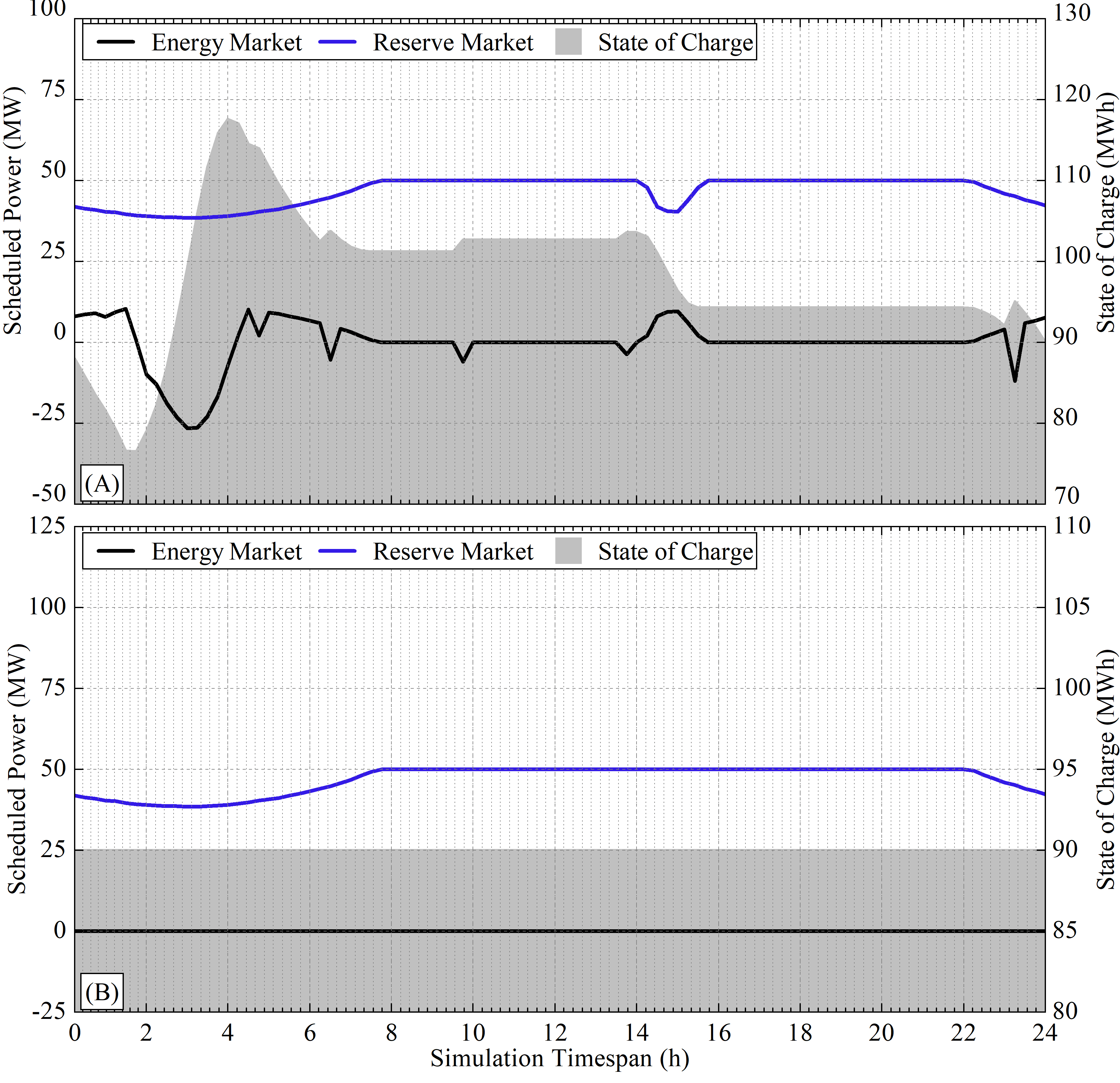}
	\caption{Scheduled power and SOC of BESS when it participates in energy and reserve markets, using (A) framework without degradation cost model; (B) framework with degradation cost model. The solid black and dashed blue curves denote BESS's scheduled power in energy and reserve markets, respectively; The grey area denotes the BESS's SOC.}
	\label{C2}
\end{figure}

This case compares the BESS's performance with and without modeling its degradation cost, when the BESS participates in both energy and reserve markets. Figs. \ref{C2}(A) and \ref{C2}(B) show the BESS's scheduled power (in MW) and SOC (in MWh) when the BESS participates in the energy and reserve markets, without and with the degradation model, respectively.

In Fig. \ref{C2}(A), the BESS prefers participating in the reserve market more than the energy market. In most hours, the BESS's scheduled power in the reserve market reaches its 50 MW charge/discharge limit. The BESS also performs inter-temporal energy arbitrage to gain revenue from the energy market, but the amount of energy arbitrage is limited as the BESS reaches its charge/discharge limit. BESSs with higher charge/discharge limits may participate more in the energy market. 

As shown in Fig. \ref{C2}(B), when the degradation cost is modeled, the reserve market becomes the only source of revenue for BESS, and BESS does not participate in the energy market. In essence, as reserve deployment is not modeled in this work, participation in the reserve market does not incur any degradation cost for the BESS. Hence, BESS provides reserve services as much as it can and does not perform any energy arbitrage between hours as the revenue generated by this activity does not overcome the degradation cost in this test system. In this case, the battery degradation does not have a significant impact on the BESS's operation in the reserve market, but this degradation reduces the BESS's energy market participation significantly.

\subsection{Case 3: Energy and Frequency Regulation Markets}
\begin{figure}[!b]
	\centering
	\includegraphics[width=3in, height=2.755in]{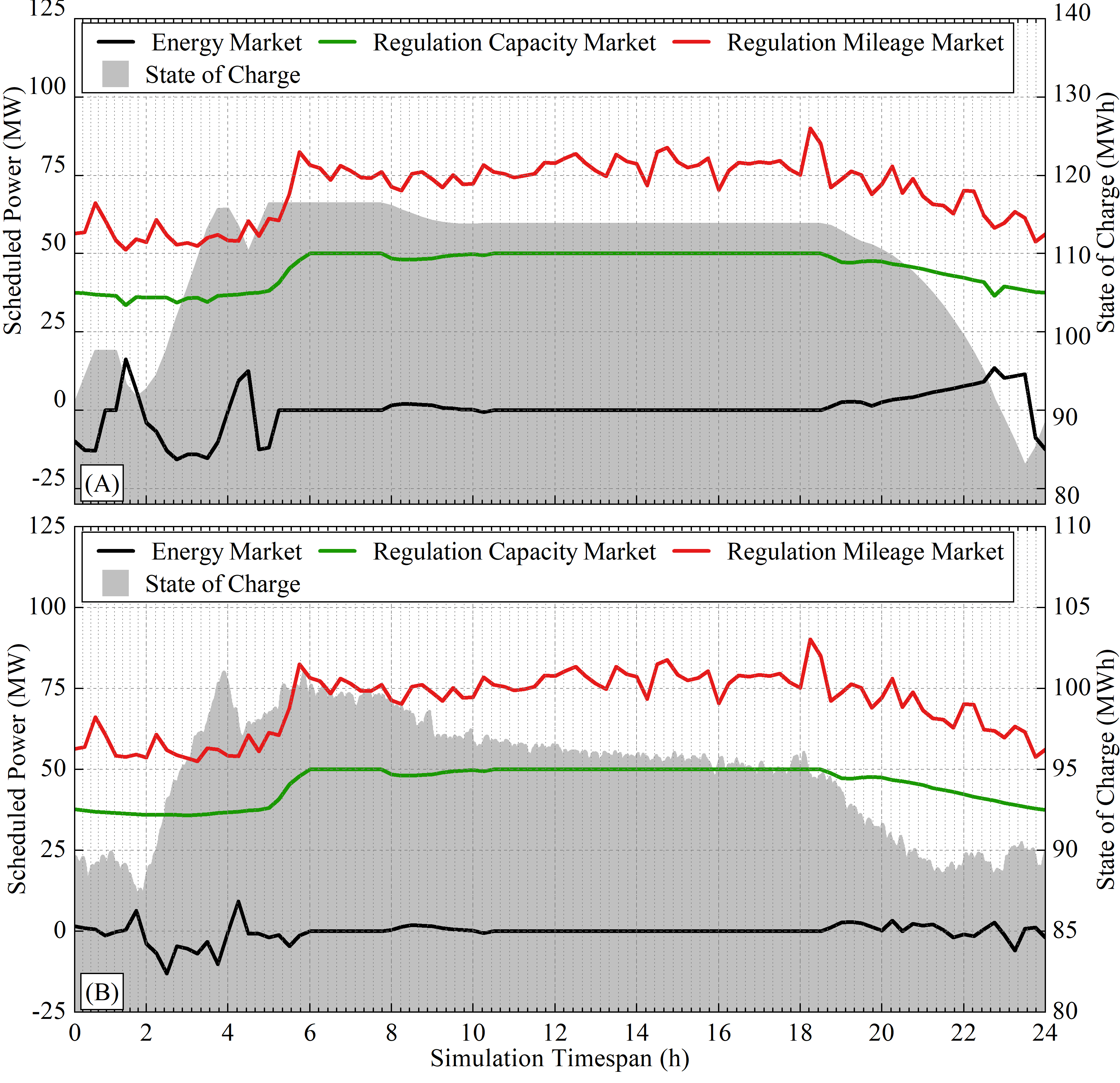}
	\caption{Scheduled power and SOC of BESS when it participates in energy and regulation markets, using (A) framework without degradation cost model; (B) framework with degradation cost model. The black, green, and red curves denote BESS's scheduled power in energy, regulation capacity, and regulation mileage markets, respectively; The grey area denotes the BESS's SOC.}
	\label{C3}
\end{figure}

This case compares the BESS's performance with and without the degradation cost model, when the BESS participates in both energy and regulation markets. Figs. \ref{C3}(A) and \ref{C3}(B) show the BESS's scheduled power (in MW) and SOC (in MWh) when the BESS participates in the energy and regulation markets, without and with the degradation model, respectively.

In this case, the simulation results without (Fig. \ref{C3}(A)) and with (Fig. \ref{C3}(B)) the degradation cost model are more similar to each other than those in the previous cases. Under both frameworks, the BESS participates the most in the frequency regulation market. In Fig. \ref{C3}(A), when battery degradation is neglected, BESS also participates in the energy market and performs energy arbitrage when its scheduled power for regulation capacity provision does not reach its charge/discharge limit at the beginning and end of the day. A BESS with a higher charge/discharge limit may participate more in the energy market. In Fig. \ref{C3}(B), when battery degradation is considered, the BESS limits its energy market participation to only compensating for the energy discharged due to following AGC signals. Although the AGC signals have zero-mean over each market clearing interval, the battery's charge/discharge efficiency is not 100\%. In the framework with models for the degradation cost and AGC signal dispatch, this fact leads to a SOC reduction when the BESS follows the AGC signals. Therefore, BESS needs to compensate for the energy discharged due to AGC signal following to maintain a similar SOC at the beginning and end of the day. 

The SOC curve of Fig. \ref{C3}(B) has small fluctuations (on top of the overall shape) all over the simulation horizon. These fluctuations are low-depth charge/discharge cycles caused by following AGC signals. As these low-depth charge/discharge cycles will result in low degradation costs, the BESS's regulation market participation will not cause significant degradation costs. Hence, degradation cost modeling in these simulations does not affect the operation pattern of BESS in the regulation market a lot, and this pattern is similar in \ref{C3}(A) and \ref{C3}(B).

%As \yellow{low-depth} charge/discharge cycles are equivalent to low degradation cost, the cost of BESS's participation in the frequency regulation market is low and worth its revenue. Hence, degradation cost modeling in these simulations does not affect the operation pattern of BESS in the regulation market a lot, and this pattern is similar in \ref{C3}(A) and \ref{C3}(B).

\subsection{Case 4: Energy, Reserve and Regulation Markets}
This case compares the BESS's performance with and without modeling its degradation cost, when the BESS participates in all the energy, reserve, and regulation markets. Figs. \ref{C2}(A) and \ref{C2}(B) show the BESS's scheduled power (in MW) and SOC (in MWh) when the BESS participates in these markets, without and with the degradation model, respectively.

\begin{figure}[!h]
	\centering
	\includegraphics[width=3in, height=2.755in]{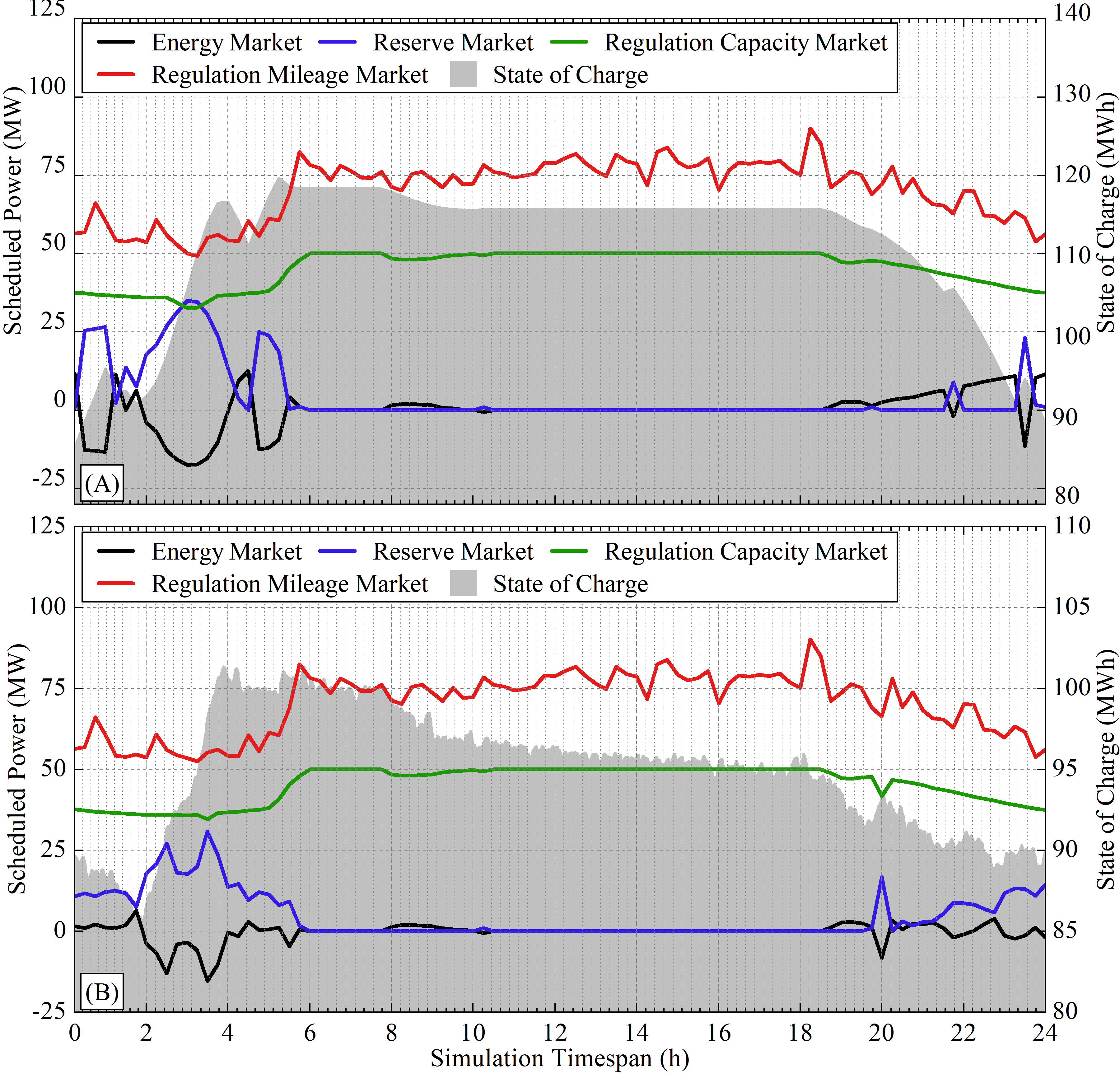}
	\caption{Scheduled power and SOC of BESS when it participates in energy, reserve, and regulation markets, using (A) framework without degradation cost model; (B) framework with degradation cost model. The black, blue, green, and red curves denote BESS's scheduled power in energy, reserve, regulation capacity, and regulation mileage markets, respectively; The grey area denotes the BESS's SOC.}
	\label{C4}
\end{figure}

Comparison of Fig. \ref{C4}(A) with Fig. \ref{C4}(B) shows that similar to the previous cases, the operation of BESS in ancillary services markets does not change a lot by considering degradation costs in this test system. This happens since 1) the reserve deployment is not modeled; and 2) providing regulation services does not cause high degradation costs. However, the BESS's energy market participation is reduced after considering degradation costs, since doing energy arbitrage in the energy market is less profitable for BESSs with degradation costs.

\subsection{Comparative Analysis for BESS's Revenue and Cost}

\begin{figure}[!h]
	\centering
	\includegraphics[width=3in, height=1.22in]{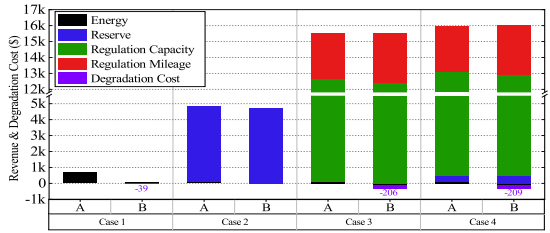}
	\caption{BESS's total revenue from each market and its degradation cost in Cases 1{\texttildelow}4, using (A) framework without degradation cost model; (B) framework with degradation cost model.}
	\label{all}
\end{figure}

Fig. \ref{all} shows the BESS's total revenue from each market and its degradation cost in Cases 1{\texttildelow}4, using frameworks with and without the degradation cost model.

Comparison of Bar A with Bar B for each case in Fig. \ref{all} validates our previous observations for this synthetic test case system build upon real-world data. First, considering batteries' degradation cost in BESS operation scheduling significantly reduces BESS's participation and revenue in the energy market. Second, BESS's participation and revenue in reserve and regulation markets do not change significantly by modeling degradation costs. However, modeling reserve deployment in the studies may affect the results for the reserve market. 

Fig. \ref{C3}(B) and Fig. \ref{C4}(B) show the degradation costs in Cases 3{\texttildelow}4 are mostly associated with the BESS's regulation market participation and AGC signal following. Fig. \ref{all} shows the total degradation costs in Cases 3{\texttildelow}4 are around 1\% of the BESS's revenue from the regulation market, which is negligible in comparison to the revenue. Therefore, neglecting the degradation costs caused by AGC signal following should not have a considerable impact on the results. This neglection could significantly reduce the optimization complexity and computation time. It could be adopted in future work when other BESS/market operational details need to be considered.

\section{Conclusion}

Based on our previously proposed frameworks for the participation of a price-maker BESS in real-time energy, reserve, and pay as performance frequency regulation markets with and without battery degradation cost model, this paper conducts comparative case studies to investigate the impact of battery degradation on BESS's revenue, cost, and operations in energy and ancillary services markets. Simulation results using synthetic test system build upon real-world data shows that 1) considering degradation cost may significantly reduce BESS's energy market participation; 2) the BESS's participation pattern in the reserve and regulation markets is not significantly impacted by the battery degradation cost. 

Future work can be focused on considering parameter uncertainties, studying other ancillary services markets, and modeling renewable resources in the frameworks.

\bibliographystyle{IEEEtran}
\bibliography{IEEEabrv,ref}

% \begin{thebibliography}{1}

% \bibitem{IEEEhowto:kopka}
% H.~Kopka and P.~W. Daly, \emph{A Guide to \LaTeX}, 3rd~ed.\hskip 1em plus
%   0.5em minus 0.4em\relax Harlow, England: Addison-Wesley, 1999.

% \end{thebibliography}

% that's all folks
\end{document}